\documentclass[]{aa}
\usepackage{graphicx}
\begin{document}

\title{An empirical temperature calibration for the $\Delta a$ 
photometric system. II. The A-type and mid F-type stars}
\author{E.~Paunzen, A.~Schnell, H.M.~Maitzen}

\mail{Ernst.Paunzen@univie.ac.at}

\institute{Institut f{\"u}r Astronomie der Universit{\"a}t Wien,
           T{\"u}rkenschanzstr. 17, A-1180 Wien, Austria}

\date{Received 23 January 2006 / Accepted 22 June 2006}
\titlerunning{An empirical temperature calibration for $\Delta a$: A and F-type stars}{}

\abstract{With the $\Delta a$ photometric system, it is possible to
study very distant galactic and even extragalactic clusters 
with a high level of accuracy. This can be done with a classical
color-magnitude diagram and appropriate isochrones. The new
calibration presented in this paper is a powerful extension.}
{For open clusters, the reddening is straightforward for
an estimation via Isochrone fitting and is needed in order 
to calculate the
reddening-free, temperature sensitive, index $(g_1-y)_0$. As a last
step, the calibration can be applied to individual stars.}
{Because no a-priori reddening-free photometric parameters are
available for the investigated spectral range, we have applied
the dereddening calibrations of the Str{\"o}mgren $uvby\beta$
system and compared them with extinction models for the Milky Way.
As expected from the sample of bright stars, the extinction is
negligible for almost all objects. As a next step,  
already established calibrations within
the Str{\"o}mgren $uvby\beta$, Geneva 7-color, and Johnson $UBV$ systems were
applied to a sample of 282 normal stars to derive
a polynomial fit of the third degree for the averaged effective temperatures to 
the individual $(g_1-y)_0$ values.}
{We derived an empirical temperature calibration for the $\Delta a$ 
photometric system for A-type to mid F-type with a 
mean of the error for the whole sample of 
$\Delta T_{eff}$ is 134\,K, which is lower than the value
in Paper I for hotter stars. No
statistically significant effect
of the rotational velocity on the precision 
of the calibration was found.}
{We have derived a new
intrinsically consistent, empirical, effective temperature calibration
for a spectral range from early B-type to mid F-type, luminosity class V to III
stars within the photometric $\Delta a$ system. The statistical
mean error over the complete spectral range of about 140 to 240\,K will allow to
individual objects of far distant galactic be studied as well as extragalactic
clusters with high accuracy.}

\maketitle

\section{Introduction}

We present the continuation of our efforts to derive an empirical 
temperature calibration for the photometric $\Delta a$ system (characteristics
described in Paunzen et al. 2005b). In the first paper (Paunzen et al.
2005d, Paper I), the results for the B-type stars and the applied methods
were published. In total, 225 stars were used to derive effective 
temperatures within the Str{\"o}mgren $uvby\beta$, Geneva 7-color, and 
Johnson $UBV$ systems based on $(u-b)$, $X$, and $(B-V)_0$, respectively. 
The final calibration for $(g_1-y)_0$ in the $\Delta a$ photometric system 
is valid for effective temperatures between 33000 and 10000\,K
and yields a statistical mean error of 238\,K for the whole spectral range.

In this paper we investigate the A-type to mid F-type objects that 
exhibit an increase line blanketing and luminosity effects
without the availability of any a-priori reddening-free parameter. Only
Str{\"om}gren $\beta$ does, in general, not depend on the extinction. 
However, it is sensitive not only to the effective temperature alone but also to the
luminosity (Gerbaldi et al. 1999).
Based on the methods used in Paper I, the following parameters were used
for our purpose: $(b-y)_0$, $(B2-V1)_0$, and $(B-V)_0$. 
The scatter of the derived effective temperatures for 
spectral types between A0 and A3 is not larger than that for later
type stars because our sample of bright galactic-field stars is almost
free of reddening.

Applying the same selection criteria as in Paper I yields
282 luminosity class V to III A-type to mid F-type stars. The derived mean
errors for the effective temperature calibrations within all photometric
systems are smaller than those from Paper I.

With the established intrinsically consistent, empirical, effective temperature calibrations for
B-type to mid F-type stars,
it is now possible to study individual objects in
very distant galactic open clusters (Paunzen et al.
2005c) and extragalactic systems (Paunzen et al. 2005a)
for which, in general, no photometric data are available within a standard system.

\begin{figure}
\begin{center}
\includegraphics[width=80mm]{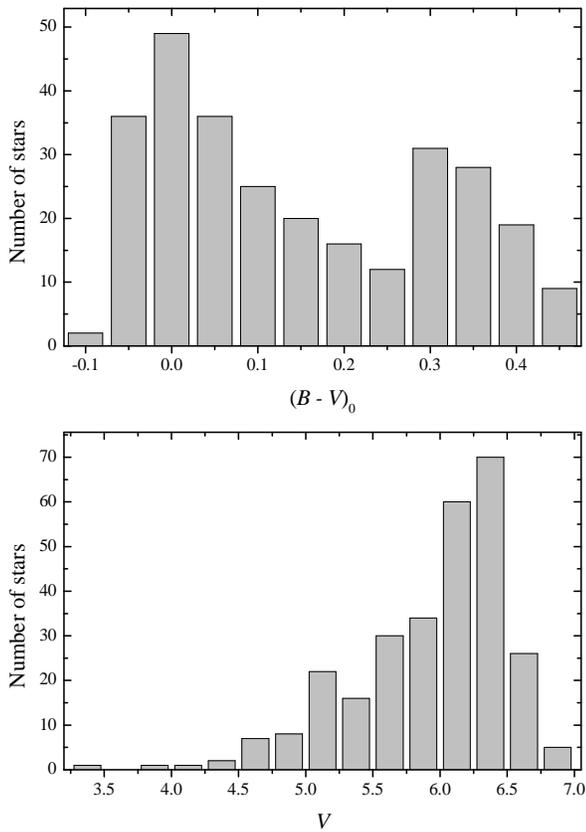}
\caption{The distribution of Johnson $(B-V)_0$ (upper panel)
and $V$ (lower panel) for
our sample of 282 main sequence A to mid F-type objects.
The distribution of $(B-V)_0$ shows two maxima at spectral types
of about A0 and F0.}
\label{hists}
\end{center}
\end{figure}

\section{Sample of program stars} \label{sops}

The selection criteria for the A-type to mid F-type objects are
exactly the same as in Paper I and will not be repeated here. 
The Johnson, Geneva and Str{\"o}mgren colors are excerpted 
from the General Catalogue of 
Photometric Data (GCPD, http://obswww.unige.ch/gcpd/gcpd.html,
Mermilliod et al. 1997). The $\Delta a$ photometry is from the paper
by Vogt et al. (1998).

The final list (Table 1) includes 282 objects and is only available in electronic 
form at the CDS or upon request from the first author. This table includes the
identification of objects, the complete Str{\"o}mgren $uvby\beta$, $(g_{1}-y)_0$,
$(B-V)_0$, and $(B2-V1)_0$ values, $V$ magnitudes, $E(b-y)$ values, 
effective temperature with the corresponding errors, 
$v\sin i$ values, and spectral types, respectively.

The distribution of Johnson $V$ and $(B-V)_0$ for our sample is shown in 
Fig. \ref{hists}. The estimation of the reddening is explained in more detail
in Sect. \ref{reddening}. The distribution of $V$ is comparable to the one 
published in Paper I, peaking at $V$\,=\,6.25,\,mag, 
whereas the one for $(B-V)_0$ exhibits two maxima at spectral types
of about A0 (+0.0\,mag) and F0 (+0.3\,mag, see also Table \ref{standard}). 
However, there is a statistically significant number of objects in the complete
investigated spectral range. 

\section{Estimation of the reddening} \label{reddening}

Our sample comprises stars later than a spectral type of A0. A typical
A0 main sequence object has a mean absolute magnitude of about +0\,mag. 
A visual magnitude of 7\,mag (Fig. \ref{hists})
then corresponds to a distance of 250\,pc, for which the reddening can 
be almost neglected in all directions (Neckel et al. 1980). 

The following photometric calibrations in the Str{\"o}mgren $uvby\beta$ system
were used to estimate the reddening according to the valid spectral range
that is estimated by the standard relations of the different indices:
\begin{itemize}
\item A0\,$-$\,A3: Crawford (1978) and Hilditch et al. (1983)
\item A3\,$-$\,F0: Crawford (1979) and Domingo \& Figueras (1999)
\item later than F0: Crawford (1973) and Schuster \& Nissen (1989)
\end{itemize}
The calibrations are
not very reliable for stars with spectral types between A0 and A3 (Gerbaldi et
al. 1999), mainly because for these stars, the reddening-free parameter $\beta$
is no longer a temperature indicator alone but is also sensitive to the
luminosity, therefore the standard relation of $(b-y)_0$ versus $m_0$ as
listed by Hilditch et al. (1983) and their method of calibration
were used. Only some objects that fall outside the
given relation were dereddened using the calibration by Crawford (1978). 

Furthermore, we used the interstellar extinction model by Chen et
al. (1998) to derive the reddening for all program stars as described in Paper I. 
The values from the calibration of the Str{\"o}mgren $uvby\beta$ and the model by
Chen et al. (1998) very closely agree. As expected,
all objects have a calibrated total absorption $A_V$ of less than 0.35\,mag
with 233 (82\% of the complete sample) stars even lower than 0.05\,mag. Taking 
the following relations into account:
\begin{eqnarray}
A_V &=& 3.1E(B-V) = 4.3E(b-y) = 4.95E(B2-V1) \nonumber \\
    &=& 7.95E(g_1-y),
\end{eqnarray}
the effect of the reddening on the calibration for our sample can be
neglected and does not introduce a significant error source.

\section{The calibration of the effective temperature} \label{tcotet}

As in Paper I, the first step was to derive the effective temperature
for each individual star within the Geneva, Str{\"o}mgren, and Johnson 
photometric systems. For this purpose we used the reddening relations
as listed in Eq. 1 to calculate the unreddened indices that are 
necessary to make the proper calibration as listed.
\\
{\it Geneva system:} the calibration by K{\"u}nzli et al (1997) is the
most recent for the investigated spectral range. For intermediate stars hotter than
8500\,K, they use the parameters $pT$ and $pG$, which are linear combinations
of the seven Geneva colors. Those two indices can be dereddened with the
following relations: $pT_0$\,=\,$pT$\,$-$\,$E(B2-V1)$ and 
$pG_0$\,=\,$pG$\,$-$\,$1.1E(B2-V1)$.
For cooler objects, the grids of $m_2$ and $d$
versus $(B2-V1)_0$ serve as a calibration tablet. 
The definition of these indices are listed in Golay (1994). \\
{\it Str{\"o}mgren system:} Napiwotzki et al. (1993) investigated
several calibrations based on $a_0$ and $r^{\ast}$ for hotter,
as well as $\beta$, and $c_0$ for cooler objects yielding a rather
unsatisfactory result. Finally, they established a $T_{eff}$ versus
$(b-y)_0$ relation (Eq. 10 therein), which was applied to our 
sample.\\
{\it Johnson system:} we used the semi-empirical $T_{eff}$ versus
$(B-V)_0$ relation as listed in Gray (1992; Eq. 15.14). It is 
based on synthetic
colors from theoretical stellar atmospheres that are normalized to 
observations of spectroscopic binary systems, as well as bright stars.
It is given as
\begin{eqnarray}
\log T_{eff} = &+&3.988 - 0.881(B-V)_0 + 2.142(B-V)_0^2 \nonumber \\
&-& 3.614(B-V)_0^3 + 3.2637(B-V)_0^4 \nonumber \\
&-& 1.4727(B-V)_0^5 + 0.2600(B-V)_0^6
\end{eqnarray}
and is valid for all A-type to mid F-type, luminosity class III to V objects.
It superseded the relation listed in Code et al. (1976).

The individual effective temperature values for the three
photometric systems were first tested
for their intrinsic consistency and then averaged. The final values,
together with the standard deviations of the means, are listed in the
Table 1 which is electronically-available. No statistical significant
outliers in any photometric system were detected.

\begin{figure}
\begin{center}
\includegraphics[width=88mm]{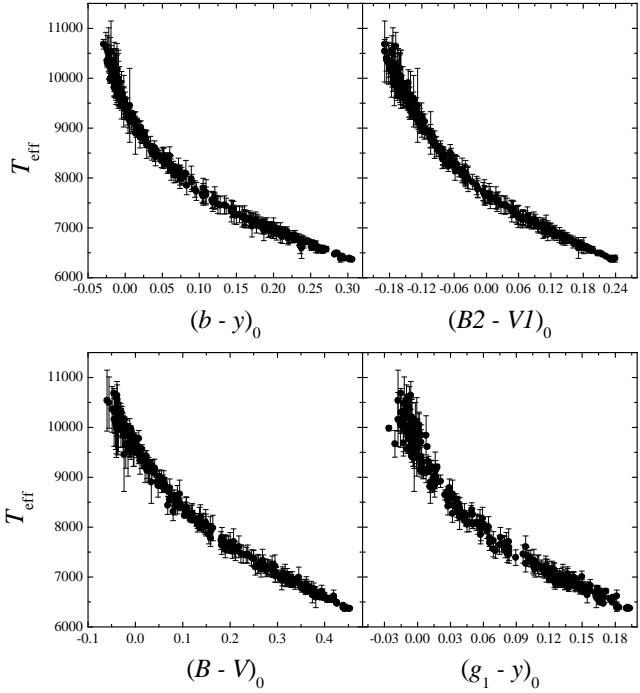}
\caption{Mean relation between the effective temperature and 
$(b-y)_0$, $(B2-V1)_0$, $(B-V)_0$, as well as $(g_1-y)_0$ for A-type
to mid F-type, luminosity class V to III objects.}
\label{teff_plot}
\end{center}
\end{figure}

\section{The result for the $\Delta a$ photometric system}

The relation between the mean effective
temperature and the different temperature sensitive indices 
for the four investigated photometric systems is shown in Fig. \ref{teff_plot}.
The errors are constant over the complete spectral range.
There is a larger scatter only 
for the Geneva $(B2-V1)_0$ index at effective temperatures hotter
than 9000\,K. We checked
this region with the grids of $pT_0$ and $pG_0$ originally used. It is
exactly where both indices become zero for the Main Sequence
and small shifts result in larger uncertainties. It therefore seems
to be an intrinsic numerical problem of the grids themselves; however
the overall statistical error of $(B2-V1)_0$ is satisfying. 
The final calibrations parameterized as third-degree polynomials
are:
\begin{eqnarray}
\log T_{eff} &=& + 3.9793(4) - 1.34(2)\cdot(b-y)_0 + \nonumber \\
&& + 4.90(15)\cdot(b-y)_0^2 - 8.06(38)\cdot(b-y)_0^3 \\
&=& + 3.8853(4) - 0.459(5)\cdot(B2-V1)_0 + \nonumber \\
&& + 1.02(2)\cdot(B2-V1)_0^2 - \nonumber \\
&& - 2.36(16)\cdot(B2-V1)_0^3 \\
&=& + 3.9825(4) - 0.670(10)\cdot(B-V)_0 + \nonumber \\
&& + 1.05(7)\cdot(B-V)_0^2 - 0.98(12)\cdot(B-V)_0^3 \\
&=& + 3.9817(8) - 1.79(5)\cdot(g_1-y)_0 + \nonumber \\
&& + 8.14(77)\cdot(g_1-y)_0^2 - 19(3)\cdot(g_1-y)_0^3
\end{eqnarray}
with the mean of the errors for the whole sample of
$\Delta T_{eff}[(b-y)_0,(B2-V1)_0,(B-V)_0,(g_1-y)_0]$
= [66,100,71,134\,K]. 
These are statistical errors for the complete sample. The errors for
individual stars are, of course, larger than that. Note that the
error for the Johnson $(B-V)_0$ calibration is surprisingly small. It 
shows the overall robustness of broad-band indices against metallicity
and luminosity effects, which makes them superior for statistical 
analysis of larger samples over a wide spectral range.
However, it has to be emphasized that it is a-priori not possible
to apply dereddening methods in the Johnson $UBV$ system for stars cooler
than B9 (Johnson 1958).

The statistical error for $(g_1-y)_0$, 134\,K is significantly smaller
than the one for the hotter stars (238\,K). But one has to keep in
mind that the absolute errors are over the complete spectral range from early
B-type to F-type constant, but only the relative one, as listed, is decreasing.

Table \ref{standard} lists the mean relations between the effective temperature and $(B-V)_0$,
$(b-y)_0$, $(B2-V1)_0$, as well as $(g_1-y)_0$, depending on spectral types (taken
from Paunzen et al. 2005b). Only standard
spectral types of the Yerkes system (Keenan 1985) are given. The $(B-V)_0$
and $(g_1-y)_0$ values for A0 are, within the error limits, identical to
those derived in Paper I (Table 1), which guarantees an intrinsic consistent
calibration from B0 to F5 for the $\Delta a$ photometric system.

As in Paper I, we checked the effect of high $v\sin i$ objects
on the derived calibrations in the same manner. This might be an issue for early A-type
objects only because the rotational velocities decrease significantly
for cooler objects (Fekel et al. 2004). Again, no
statistical significant effect of the rotational velocity,
which can be distinguished from other error sources, was found on the precision 
of the calibration. 

The detailed calibration procedure for determining  the effective
temperature within the $\Delta a$ photometric system is given in Paper I.
We only give a short overview of the main method here. After the
estimation of the reddening $E(g_1-y)$, the dereddened standard $(g_1-y)_0$ has
to be calculated and the calibration for the appropriate spectral region
applied. As a final check, a comparison with the values for other
photometric systems, if available, should be performed.

\addtocounter{table}{1}
\begin{table}[t]
\begin{center}
\caption{Mean relation between the effective temperature and $(B-V)_0$,
$(b-y)_0$, $(B2-V1)_0$, as well as $(g_1-y)_0$ for A-type to mid F-type, 
luminosity class V to III objects. Only spectral types according to the
Yerkes system (Keenan 1985) are listed.}
\label{standard}
{\scriptsize
\begin{tabular}{crcccc}
\hline
\hline
Spec. & $T_{eff}$ & $(B-V)_0$ & $(b-y)_0$ & $(B2-V1)_0$ & $(g_1-y)_0$ \\
\hline
A0 & 10000 & $-$0.025 & $-$0.015 & $-$0.160 & $-$0.010 \\
A2 &  8750 & +0.067   & +0.031   & $-$0.093 & +0.025   \\
A3 &  8300 & +0.112   & +0.055   & $-$0.060 & +0.042   \\
A5 &  7900 & +0.161   & +0.082   & $-$0.024 & +0.062   \\
A7 &  7500 & +0.221   & +0.120   & +0.022   & +0.087   \\
F0 &  7050 & +0.304   & +0.185   & +0.087   & +0.125   \\
F2 &  6750 & +0.368   & +0.238   & +0.141   & +0.155   \\
F3 &  6650 & +0.390   & +0.254   & +0.159   & +0.165   \\
F5 &  6450 & +0.433   & +0.284   & +0.196   & +0.184   \\
\hline
\end{tabular}
}
\end{center}
\end{table}

\section{Conclusion}

In the second and last parts of this series, we established 
an empirically effective temperature calibration for the $\Delta a$
photometric system for main sequence (luminosity class V to III)
A-type to mid F-type star. Applying the same methods as
in Paper I, the calibrations in the Str{\"o}mgren $uvby\beta$, 
Geneva 7-color, and Johnson $UBV$ photometric systems
based on $(b-y)_0$, $(B2-V1)_0$, and $(B-V)_0$, respectively are
used to derive effective temperature for 282 normal type 
objects. The final calibrations are expressed third-degree 
polynomials. 

As expected from the brightness and thus the distance from the
Sun ($<$\,250\,pc) for our sample, the reddening can almost be
neglected for most of the objects. This is important because
no reddening-free photometric parameter is a-priori available 
for the whole investigated spectral domain.

The established standard system guarantees a direct link to 
the calibration for B-type stars published in Paper I. It is
therefore possible to calibrate effective temperatures via 
$(g_1-y)_0$ in a consistent way for spectral types from B0 to
F5 for luminosity classes V to III with a constant absolute
error of a few percent.

This will allow us to independently investigate individual stars of 
distant galactic, as well as extragalactic open clusters within the $\Delta a$
photometric system, to a high level of accuracy.

\begin{acknowledgements}
This research was performed within the projects  
{\sl P17580} and {\sl P17920} of the Austrian Fonds zur F{\"o}rderung der 
wissen\-schaft\-lichen Forschung (FwF). 
Use was made of the SIMBAD database, operated at the CDS, Strasbourg, France, and
of the NASA's Astrophysics Data System. We would like thank the referee for
her/his important comments pointing at serious problems in the first version. 
\end{acknowledgements}


\begin{thebibliography}{}
\bibitem[]{} Chen, B., Vergely, J. L., Valette, B., Carraro, G. 1998, A\&A, 336, 137
\bibitem[]{} Code, A. D., Bless, R. C., Davis, J., Brown, R. H. 1976, ApJ, 203, 417
\bibitem[]{} Crawford, D. L. 1975, AJ, 80, 955
\bibitem[]{} Crawford, D. L. 1978, AJ, 83, 48
\bibitem[]{} Crawford, D. L. 1979, AJ, 84, 1858
\bibitem[]{} Domingo, A., \& Figueras, F. 1999, A\&A, 343, 446
\bibitem[]{} Fekel, F. C., Warner, P. B., \& Kaye, A. B. 2004, in 
Stellar Rotation, ed. A. Maeder, \& Ph. Eenens, 
San Francisco, ASP Conf. Series, 215, 53
\bibitem[]{} Gerbaldi, M., Faraggiana, R., Burnage, R., Delmas, F., G{\'o}mez, A. E., Grenier, S.
1999, A\&AS, 137, 273
\bibitem[]{} Golay, M. 1994, in The MK process at 50 years. A powerful tool for 
astrophysical insight, ed. Ch. Corbally, R. O. Gray, \& R. F. Garrison,
San Francisco, ASP Conf. Series, 60, 164
\bibitem[]{} Gray, D. F. 1992, The Observation and Analysis of Stellar Photospheres,
Cambridge University Press
\bibitem[]{} Hilditch, R. W., Hill, G., \& Barnes, J. V., 1983, MNRAS, 204, 241
\bibitem[]{} Johnson, H. L. 1958, Lowell Obs. Bull., No. 4, 37
\bibitem[]{} Keenan, P.C. 1985, in Calibration of fundamental stellar quantities,
Dordrecht, D. Reidel Publishing Co., p. 121
\bibitem[]{} K{\"u}nzli, M., North, P., Kurucz, R. L., Nicolet, B. 1997, A\&AS, 122, 51
\bibitem[]{} Mermilliod, J.-C., Mermilliod, M., \& Hauck, B. 1997, A\&AS, 124, 349
\bibitem[]{} Napiwotzki, R., Sch{\"o}nberner, D., \& Wenske, V. 1993, A\&A, 268, 653
\bibitem[]{} Neckel, Th., Klare, G., \& Sarcander, M. 1980, A\&AS, 42, 251
\bibitem[]{} Paunzen, E., Pintado, O. I., Maitzen, H. M., Claret, A., 2005a, MNRAS,
362, 1025
\bibitem[]{} Paunzen, E., St{\"u}tz, Ch., \& Maitzen, H. M. 2005b, A\&A, 441, 631
\bibitem[]{} Paunzen, E., Netopil, M., Iliev, I. Kh., Maitzen, H. M., Claret, A., Pintado, O. I.
2005c, A\&A, 443, 157
\bibitem[]{} Paunzen, E., Schnell, A., \& Maitzen, H. M. 2005d, A\&A, 444, 941 (Paper I)
\bibitem[]{} Schuster, W. J., \& Nissen, P. E. 1989, A\&A, 221, 65
\bibitem[]{} Vogt, N., Kerschbaum, F., Maitzen, H. M., Fa{\'u}ndez-Abans, M. 1998, 
A\&AS, 130, 455 
\end{thebibliography}
\end{document}